\documentclass[twocolumn,showpacs,preprintnumbers,amsmath, amssymb, prb, aps, superscriptaddress]{revtex4-1}
 \usepackage{graphicx}
 \usepackage{dcolumn}
 \usepackage{bm}
 \usepackage{color,ulem} 
\pdfoutput=1
 \begin{document}
  \title{Opening stop-gaps in plasmonic crystals by tuning the radiative coupling of surface plasmons to diffracted orders }

   \author{S. R. K. Rodriguez}\email{s.rodriguez@amolf.nl}
   \affiliation{Center for Nanophotonics, FOM Institute AMOLF, c/o Philips Research Laboratories, High Tech Campus 4, 5656 AE Eindhoven, The Netherlands}

   \author{O. T. A. Janssen}
   \affiliation{Optics Research Group, Delft University of Technology, 2628 CJ Delft, The Netherlands}

   \author{A. Abass}
   \affiliation{Department of Electronic and Information Systems (ELIS), Ghent University, Sint-Pietersnieuwstraat 41, B-9000 Ghent, Belgium}

  \author{B. Maes}
  \affiliation{Micro- and Nanophotonic Materials Group, Institut de Physique, University of Mons, Place du Parc 20, B-7000 Mons, Belgium}
  \affiliation{Photonics Research Group (INTEC), Ghent University-imec, Sint-Pietersnieuwstraat 41, B-9000 Ghent, Belgium}

  \author{G. Vecchi}
  \affiliation{Center for Nanophotonics, FOM Institute AMOLF, c/o Philips Research Laboratories, High Tech Campus 4, 5656 AE Eindhoven, The Netherlands}

  \author{J. G\'{o}mez Rivas}
  \affiliation{Center for Nanophotonics, FOM Institute AMOLF, c/o Philips Research Laboratories, High Tech Campus 4, 5656 AE Eindhoven, The Netherlands}
  \affiliation{COBRA Research Institute, Eindhoven University of Technology, P.O. Box 513, 5600 MB Eindhoven, The Netherlands}

\date{\today}

\begin{abstract}
By tuning the radiative coupling of localized surface plasmons to
diffracted orders, we demonstrate how stop-gaps in plasmonic
crystals of nanorods may be opened and tuned. The stop-gap arises
from the mutual coupling of surface lattice resonances (SLRs), which
are collective resonances associated with counter-propagating
surface polaritons. We present experimental results for three
different nanorod arrays, where we show how the dispersion of SLRs
can be controlled by modifying the size of the rods. Combining
experiments with numerical simulations, we show how the properties
of the stop-gap can be tailored by tuning a single structural
factor. We find that the central frequency of the stop-gap falls
quadratically, the frequency width of the stop-gap rises linearly,
and the in-plane momentum width of the standing waves rises
quadratically, as the width of the nanorods increases. These
relationships hold for a broad range of nanorod widths, including
duty cycles of the array between 20\% to 80\%. We discuss the
physics in terms of a coupled oscillator analog, which relates the
tunability of the stop-gaps to the coupling strength of plasmonic
modes.
\end{abstract}

\pacs{42.70.Qs, 78.67.Qa, 73.20.Mf, 42.25.Fx} 
\maketitle

Photonic bandgaps, i.e., frequency regions within which optical
propagation in a photonic crystal is forbidden, enable to control
the flow of light. Gaps arise in the dispersion of photonic modes as
a consequence of the optical periodicity, and the width of the gap
is determined by the degree of field concentration in the different
dielectric regions~\cite{Joann}. A more complex scenario arises in
plasmonic crystals, where light-matter interactions are no longer
dominated by the optical periodicity alone. With the penetration of
the electromagnetic field into the metal and the excitation of
surface plasmon modes, fascinating phenomena fluorish in plasmonic
crystals.  For this reason, the coupling of surface modes in
periodic metallic nanostructures has attracted much interest since
early investigations~\cite{Barnes96b}, especially for its connection
with frequency stop-gaps~\cite{Barnes96}. Coupled surface modes have
been observed in metallic gratings~\cite{Lochbihler94, Yoon},
subwavelength hole arrays~\cite{Lalanne08, Billaudeau}, nanoslit
arrays~\cite{Ropers}, and particle arrays coupled to waveguide
modes~\cite{Ghoshal&Kik09}. Despite the numerous structures that
have been investigated, very few studies have discussed the physical
origin of these gaps. One notable exception is the work by Barnes
and co-workers~\cite{Barnes96}, where the origin of the gap in
metallic sinusoidal gratings is discussed. However, the analysis
therein contained is not easily extended to more complex plasmonic
structures, which renders difficult the emergence of a simple,
intuitive explanation on the origin of the gap.

It is the aim of this work to provide an intuitive framework in
which the opening of frequency stop-gaps in plasmonic crystals can
be understood in terms of the coupling strength of the surface modes
involved. We investigate nanorod arrays supporting Surface Lattice
Resonances (SLRs), which are dispersive and spectrally narrow
collective resonances arising from the diffractive coupling of
localized surface plasmons~\cite{Auguie&Barnes08, Crozier,
Kravets08, Vecchi09b, Bitzer}. This coupling occurs near the
condition at which a diffraction order changes from radiating to
evanescent, i.e. at the Rayleigh anomaly.  In a recent work, we
discussed the coupling of bright and dark SLRs in nanorod
arrays~\cite{Rodriguez11}. The associated stop-gap, modal
symmetries, and the very high quality factors of SLRs were therein
discussed. In this paper, we demonstrate how stop-gaps associated
with coupled SLRs can be selectively opened by tuning the radiative
coupling of localized surface plasmons to diffracted orders. We
present experimental results for three different arrays with varying
nanorod dimensions but equal lattice constants. By combining
experiments with numerical simulations, we elucidate the influence
of the width of the nanorod on the dispersion of SLRs. We also find
scaling laws for the properties of the gap as a function of the
nanorod width. As we will show, these scaling laws are related to
the coupling strength of the surface modes involved, and they are
valid for a wide range of duty cycles.

\begin{figure*}
\centerline{\scalebox{0.8}{\includegraphics{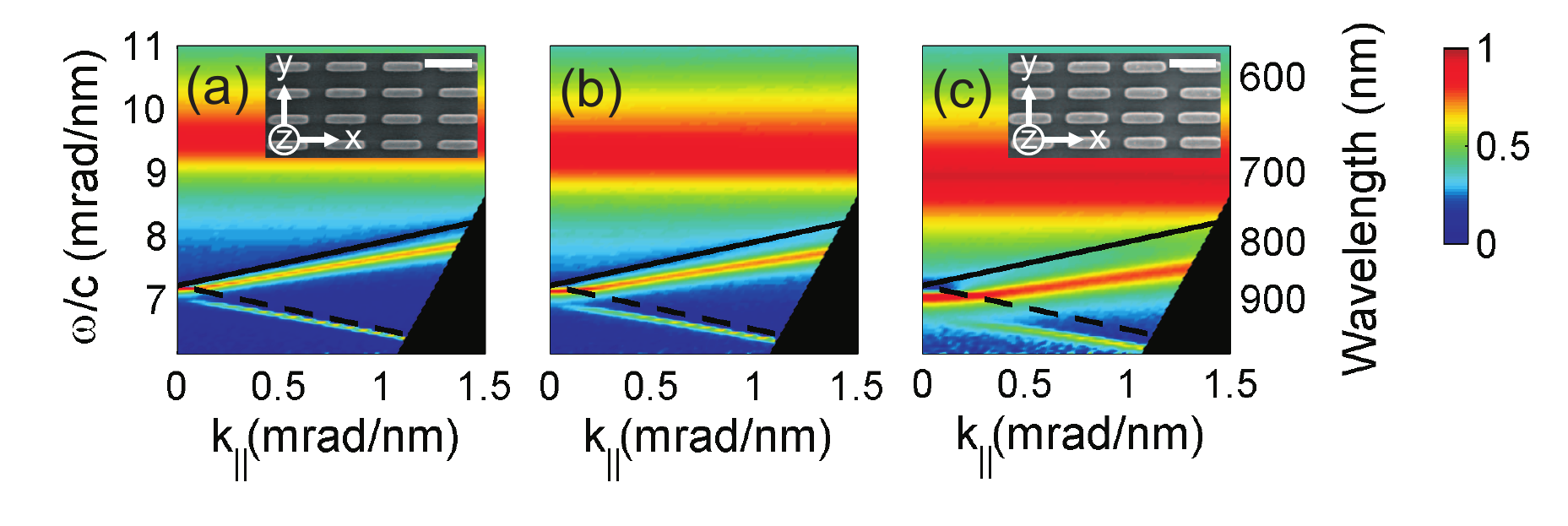}}}
\caption{(Color online) Experimental extinction spectra as a
function of the wave vector component parallel to the surface of the
array. Scanning electron microscope images of the nanorod arrays
yielding the spectra in Figures (a) and (c) are shown in the
corresponding insets; the scale bar denotes 500 nm. The black solid
and dashed lines indicate the $(+1,0)$ and $(-1,0)$ Rayleigh
anomalies, respectively. The incident light is polarized parallel to
the width of the nanorods (y-direction), which is (a) 85 nm, (b)
95nm, and (c) 115 nm. The broad, dispersionless resonance on the
high frequency side of each spectrum is the dipolar localized
surface plasmon resonance along the width of the nanorods. The
narrower and dispersive resonances below the Rayleigh anomalies are
the surface lattice resonances.}\label{gaps_fig1}
\end{figure*}
We have investigated experimentally three $3 \times 3$ mm$^2$ arrays
of gold nanorods fabricated on a silica substrate using Substrate
Conformal Imprint Lithography (SCIL)~\cite{scil}. The arrays were
embedded in a uniform surrounding medium by placing a silica
superstrate preceded by n=1.45 index matching fluid to ensure good
optical contact.  The three arrays have lattice constants $a_x=600$
nm and $a_y=300$ nm, but they comprise nanorods which differ in
size. The nanorods have an approximately rectangular shape in the
plane of the array, and a height $= 38 \pm 2$ nm. The rod size
(length $\times$ width) was tuned by varying the exposure dose of
the electron beam when preparing the master for nanoimprint. This
procedure yielded rods of size (a)$420 \times 85 $ nm$^2$, (b)$430
\times 95 $ nm$^2$, and (c)$450 \times 115 $ nm$^2$, which
correspond to the measurements in Figures~\ref{gaps_fig1}(a),
~\ref{gaps_fig1}(b), and~\ref{gaps_fig1}(c), respectively. The
tolerances of these in-plane dimensions are on the order of $\pm 10$
nm. Figures~\ref{gaps_fig1}(a) and~\ref{gaps_fig1}(c) show a
Scanning Electron Microscope (SEM) image of the corresponding array,
and a cartesian triad in the inset which we use to describe the
measurements next.

 Figure~\ref{gaps_fig1} shows the extinction,
defined as $1-T$ with $T$ the zeroth order transmittance, for the
three arrays described above. The extinction is displayed as a
function of the reduced frequency, i.e., the angular frequency
normalized by the speed of light in vacuum, and the component of the
incident wave vector parallel to the surface, which is given by
$k_\|=\frac{\omega}{c} n \sin(\theta) \hat{x}$, with $n$ the
refractive index of silica and $\theta$ the angle of incidence. The
sample was rotated around the y-axis while the y-polarized
collimated beam from a halogen lamp impinged onto the sample,
probing the short axis of the nanorods. The broad, dispersionless
extinction peak seen on the high frequency side of the spectra for
all three arrays corresponds to the excitation of Localized Surface
Plasmon Resonances (LSPRs) in the individual nanorods. The black
solid and dashed lines indicate the (+1,0) and (-1,0) Rayleigh
anomalies of the arrays, respectively. The Rayleigh anomalies are
solutions to the equation $k_{out}^2 = k_{in}^2$ sin$^2(\theta) +
m_1^2 (2 \pi / a_x)^2 + m_2^2 (2 \pi / a_y)^2 + 2 k_{in}$
sin$(\theta) m_1(2 \pi / a_x)$, where $k_{in}$ and $k_{out}$ are the
modulus of the incident and scattered wave vectors and $m_j$
($j=1,2$) are the integers defining the order of diffraction.
Physically, these so-called anomalies represent the frequency and
wave vector for which the corresponding diffracted orders are
propagating grazing to the surface of the array. The coupling of
LSPRs to the Rayleigh anomalies gives rise to the SLRs, which
manifest in the measurements as narrow and dispersive peaks in
extinction at slightly lower frequencies than the associated
Rayleigh anomalies.

 Figures~\ref{gaps_fig1}(a)-(c) show the gradual opening of a stop-gap
in the dispersion relation of SLRs as the nanorod width increases.
The gap is centered near 7 mrad/nm in Fig.~\ref{gaps_fig1}(a), but
its central frequency is lowered and its width $\Delta \omega_{gap}$
increases as the nanorods become wider. We note that this is not a
complete photonic band-gap, since it only exists for light polarized
parallel to the short axis of the nanorods and with an in-plane wave
vector component parallel to the long axis of the nanorods. For
light polarized parallel to the long axis of the nanorods, the
dipolar LSPR lies at lower energies than the ($\pm$1, 0) diffraction
orders at normal incidence, which results in a  weak diffractive
coupling~\cite{Auguie&Barnes08}. On the other hand, for an in-plane
wave vector component parallel to the short axis of the nanorods,
the ($\pm$1,0) Rayleigh anomalies are degenerate, leading to
degenerate ($\pm$1,0) SLRs and therefore no gap~\cite{Giannini10}.

Inspired by previous work explaining electromagnetic resonance
phenomena in terms of coupled oscillators~\cite{Alzar05, Halas10},
we recently introduced an analog to the plasmonic crystal consisting
of three mutually coupled harmonic oscillators~\cite{Rodriguez11}.
In this analogy, the conduction electrons in the nanorod driven by
the electromagnetic field are modeled as oscillator 1 driven by a
harmonic force $F = F_0 e^{-i\omega_s t}$. The (+1,0) and (-1,0)
Rayleigh anomalies are modeled by oscillators 2 and 3, respectively.
The equations of motion of the coupled system are,
 \begin{align}
 \ddot{x}_1 + \gamma_1 \dot{x}_1 + \omega_1^2 x_1 - \Omega_{12}^2 x_2  - \Omega_{13}^2 x_3 &= F, \nonumber \\
 \ddot{x}_2 + \gamma_2 \dot{x}_2 + \omega_2^2 x_2 - \Omega_{12}^2 x_1 - \Omega_{23}^2 x_3 &= 0, \\
 \ddot{x}_3 + \gamma_3 \dot{x}_3 + \omega_3^2 x_3  - \Omega_{13}^2 x_1 - \Omega_{23}^2 x_2 &= 0, \nonumber
 \end{align}
 with $x_j$, $\gamma_j$ , and $\omega_j$ ($j = 1,2,3$) the displacement from
equilibrium position, damping, and eigenfrequency associated with
the $j^{th}$ oscillator, respectively, and $\Omega_{jk}$ ($k=1,2,3$
and $j\neq k$) the coupling frequency between the $j^{th}$ and
$k^{th}$ oscillator. The absorbed mechanical power by oscillator 1
from the driving force, given by $P(t) = F \dot{x}_1$, is
representative of the extinct optical power by the array.
Integrating $P(t)$ over one period of oscillation and scanning the
driving frequency $\omega_s$ yields an absorbed power spectrum,
which we compare to the extinction measurements.

\begin{figure}
\centerline{\scalebox{0.6}{\includegraphics{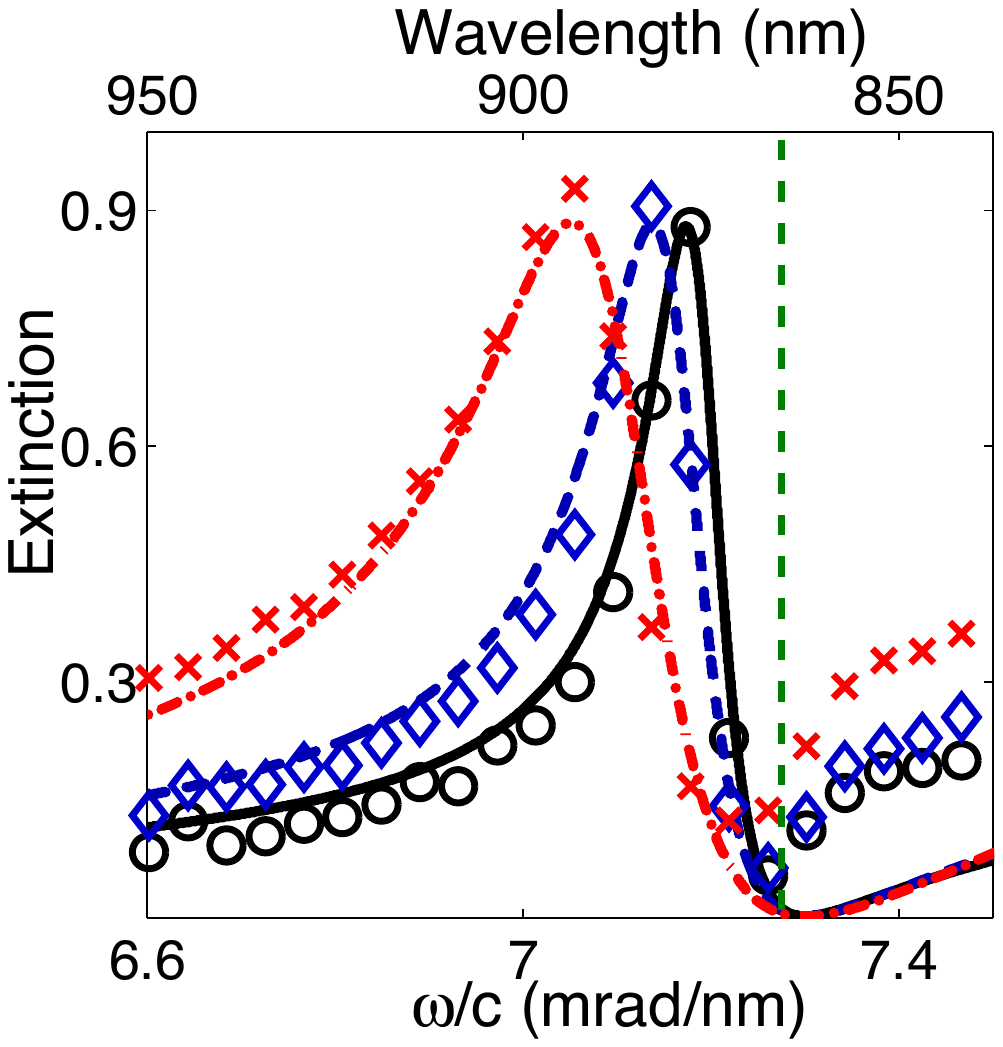}}}
\caption{(Color online) The black open circles, blue open diamonds,
and red crosses, are the experimental extinction spectra at normal
incidence for arrays of nanorods of width  85 nm,  95nm,  and 115
nm, respectively. The black solid, blue dashed, and red dash-dot
curves are calculations of the absorbed power in the coupled
oscillator model described in the text and with fitting parameters
as given in Figure 3(c). The dashed vertical line indicates the
eigenfrequency $\omega_2 = 7.3$ mrad/nm representing the Rayleigh
anomaly in the oscillator model. }\label{gaps_fig2}
\end{figure}

Figure~\ref{gaps_fig2} displays the extinction spectra for the three
arrays at normal incidence, together with calculations done with the
coupled oscillator model as fits to the measurements. Only the
(+1,0) SLR appears in the spectra, so the model reduces to two
oscillators, i.e., $\Omega_{13} = \Omega_{23} = 0$. The third
oscillator is uncoupled because the (-1,0) SLR is a dark state at
normal incidence. This condition arises due to an antisymmetric
(quadrupolar) character of the (-1,0) modal fields, which results in
a vanishing extinction  at normal incidence~\cite{Rodriguez11}.
Figure~\ref{gaps_fig1} shows that for an increasing nanorod width
the LSPR displays a diminishing center frequency and a broader
linewidth. This behavior is associated with increased retardation
and radiative damping~\cite{Maier}, which we model in the
calculations of Fig.~\ref{gaps_fig2} by lowering $\omega_1$ and
increasing $\gamma_1$. The eigenfrequency of oscillator 2,
representing the Rayleigh anomaly frequency, is set to $\omega_2 =
7.3$ mrad/nm for all three cases. In order to unambiguously
demonstrate the critical role of the coupling between LSPRs and the
Rayleigh anomaly in determining the SLR lineshape, we set the
damping of oscillator 2 (whose coupling gives rise to the SLR) equal
in all three cases, given by $\gamma_2 = 0.001$ mrad/nm. Considering
that the losses are expected to increase for the bigger nanorods,
this is unlikely the exact case in the experiments. However, as we
show next, fixing $\gamma_2 $ allows us to see the effect of
changing the coupling frequency $\Omega_{12}$, which stands for the
radiative coupling strength between the LSPR and the Rayleigh
anomaly.

\begin{figure}
\centerline{\scalebox{0.5}{\includegraphics{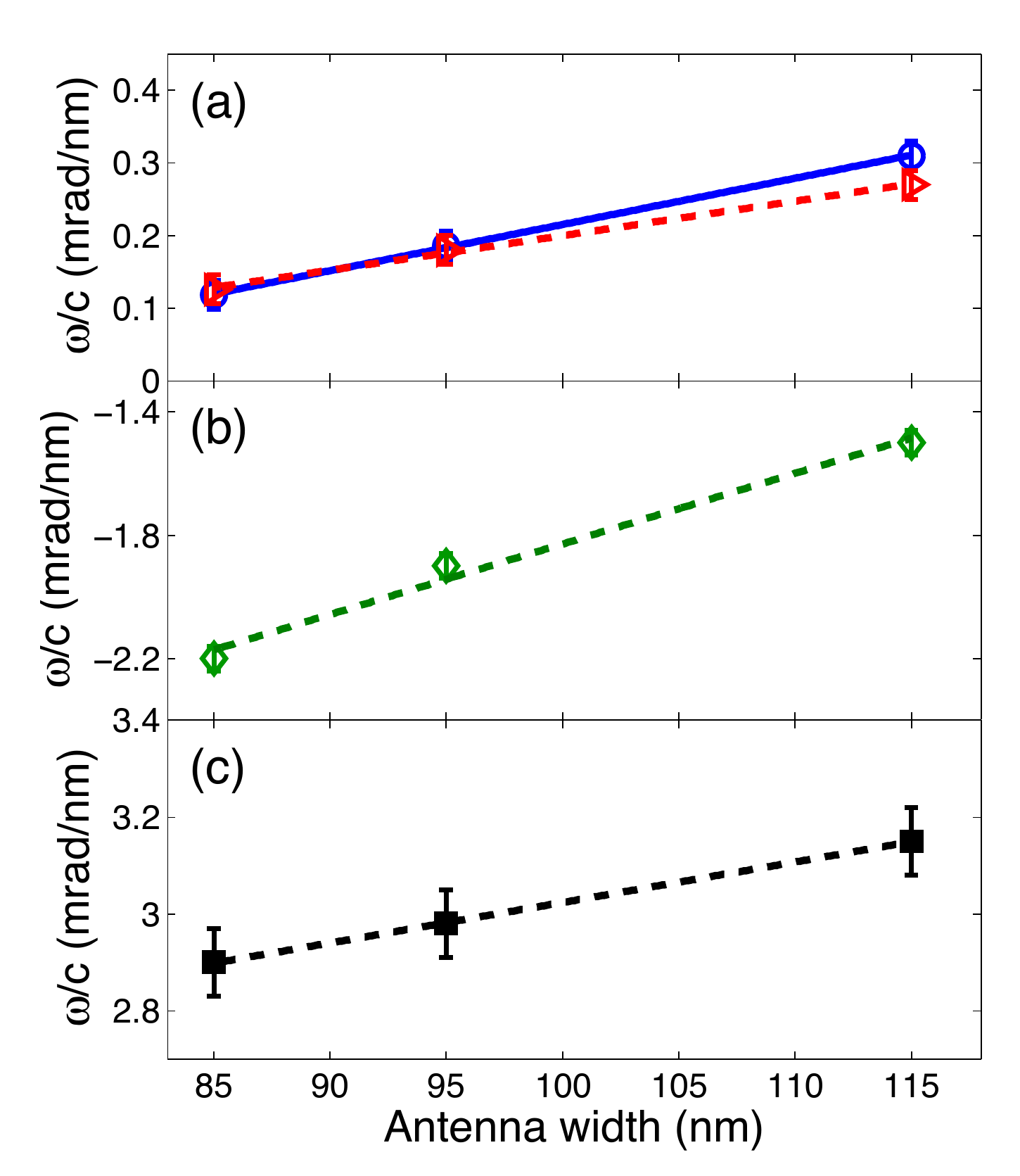}}}
\caption{(Color online) Figure (a) shows the (+1,0) SLR linewidth at
Full-Width Half Maximum (FWHM)  as blue open circles, and the
detuning of the SLR from the Rayleigh anomaly, $\omega_{RA} -
\omega_{SLR}$, as red open triangles. Both quantities are taken from
the extinction measurements at normal incidence. The solid and
dashed lines are guides to eye for the FWHM and detuning,
respectively. Figure (b) shows the frequency difference between the
Rayleigh anomaly and the Localized Surface Plasmon Resonance (LSPR),
$\omega_{RA} - \omega_{LSPR}$, as green open diamonds. The green
dashed line is a guide to eye. Figure (c) shows the coupling
frequency $\Omega_{12}$ used in the coupled oscillator model to
calculate the spectra shown in Figure 2. The black dashed curve is a
quadratic fit to $\Omega_{12}$. All quantities in (a) and (b) are
plotted as a function of the nanorod width. The error bars in (a)
and (b) stem from the resolution in the spectrometer, whereas the
error bars in (c) represent the uncertainty in the
fitting.}\label{gaps_fig3}
\end{figure}

The measurements in Fig.~\ref{gaps_fig2} show two main effects on
the SLR lineshape as the width of the nanorods increases: i) the
peak resonance frequency is increasingly detuned from the Rayleigh
anomaly,  and ii) the linewidth broadens. As it is shown next, this
behavior is explained by the change of the coupling constant between
the LSPR and the Rayleigh anomaly. In Fig.~\ref{gaps_fig3}(a) we
plot the detuning of the SLR from the Rayleigh anomaly, $\omega_{RA}
- \omega_{SLR}$, and the linewidth at Full-Width Half Maximum (FWHM)
of the SLR as a function of the width of the nanorods. It is
remarkable that these two quantities are in quantitative agreement,
which indicates a direct connection between radiative losses and the
peak resonance frequency with respect to the corresponding Rayleigh
anomaly.  The implications of this connection for sensing small
changes to the bulk refractive index by means of plasmonic
nanoparticle arrays have been recently discussed~\cite{Offermans}. A
universal scaling of the figure of merit of plasmonic sensors, which
is a function of the detuning $\omega_{RA} - \omega_{SLR}$ alone,
was therein found. In Fig.~\ref{gaps_fig3}(b) we plot $\omega_{RA} -
\omega_{LSPR}$ as a function of the nanorod width, which displays
the diminishing frequency difference between the LSPR and the
Rayleigh anomaly. This decrease in the magnitude of $\omega_{RA} -
\omega_{LSPR}$ with a simultaneous broadening of the LSPR promotes a
stronger radiative coupling of localized surface plasmons to
diffracted orders, i.e., an increase in $\Omega_{12}$.
Figure~\ref{gaps_fig3}(c) shows the values of $\Omega_{12}$ used to
fit the measurements in Fig.~\ref{gaps_fig2}.  Although in the
fitted range $\Omega_{12}$ may well be described by a linear
function, we have used a quadratic function for a reason that will
be clarified further in the text. We see that increasing
$\Omega_{12}$ detunes the SLR from the Rayleigh anomaly and broadens
it in the right amount to have an excellent agreement with the
measurements. A crucial understanding in attributing the observed
behavior to $\Omega_{12}$ mainly lies in the fact that changing
$\gamma_2$ (the intrinsic damping of oscillator 2) can not lead to
the right detuning $\omega_{RA} - \omega_{SLR}$.  We have verified
this through several calculations (not shown here) in which
$\gamma_2$ and $\Omega_{12}$ were varied independently and/or
simultaneously. Next, we consider the more general case of inclined
incidence light, where the three oscillators are mutually coupled.

\begin{figure}
\centerline{\scalebox{0.5}{\includegraphics{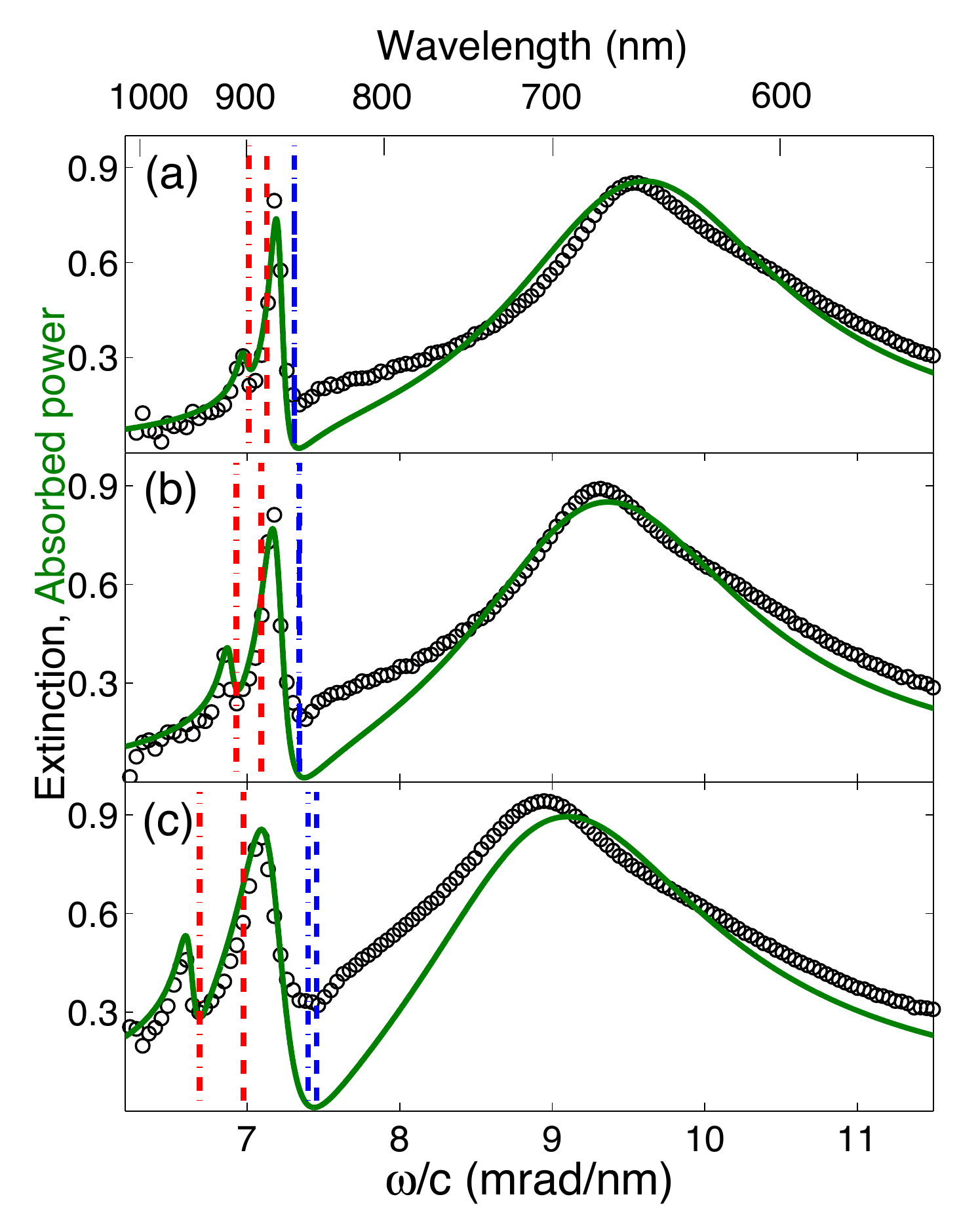}}}
\caption{(Color online) The black open circles are cuts of the
measured extinction spectra for the three arrays in Figure 1,
following the same labeling convention for the antenna width, i.e.,
(a) 85 nm, (b) 95 nm,  and (c) 115 nm. For each case the extinction
is shown at the slowdown value of $k_\|$, which is at (a) $k_\| =
0.13$ mrad/nm, (b) $k_\|= 0.18$ mrad/nm, and (c) $k_\| = 0.35$
mrad/nm. The green solid curves represent the absorbed power in the
coupled oscillator model described in the text, with coupling and
damping frequencies as given in Table 1. The blue and red dashed
lines are the (+1,0) and (-1,0) Rayleigh anomalies as predicted by
the theory (conservation of the parallel component of the wave
vector across the interface), respectively. The blue and red
dash-dot lines are the eigenfrequencies $\omega_2$ and $\omega_3$,
respectively, set in the oscillator model to calculate the absorbed
power spectrum for each case. }\label{gaps_fig4}
\end{figure}

Inspection of Fig.~\ref{gaps_fig1} points towards the connection
between the formation of standing waves in the high frequency SLR
band and the opening of the gap. We consider the gap to open at the
slowdown point for the high frequency band, i.e., where the
dispersion of the (+1,0) SLR flattens and the group velocity slows
down to zero. Figure~\ref{gaps_fig4} shows the extinction of the
three arrays at the slowdown point, which occurs at (a) $k_\| =
0.13$ mrad/nm, (b) $k_\| = 0.18$ mrad/nm, and (c) $k_\| = 0.35$
mrad/nm. In the same Figure we plot the absorbed power spectra for
the oscillator model fitting each measurement. From the frequency
and linewidth of the dispersionless LSPR we set the eigenfrequency
and damping of oscillator 1 to (a) $\omega_1 = 9.5$ mrad/nm,
$\gamma_1 = 2.3$ mrad/nm, (b) $\omega_1 = 9.2$ mrad/nm,  $\gamma_1 =
2.4$ mrad/nm, and (c) $\omega_1 = 8.8$ mrad/nm,  $\gamma_1 = 2.7$
mrad/nm. The coupling and damping frequencies used to fit the SLR
lineshapes are given in Table 1, together with the eigenfrequencies
$\omega_2$ and $\omega_3$ representing the Rayleigh anomalies. The
blue and red dashed vertical lines in Fig.~\ref{gaps_fig4} indicate
the (+1,0) and (-1,0) Rayleigh anomalies, respectively, as predicted
by the equation describing the conservation of the wave vector
component parallel to the surface of the grating. These are the
frequencies at which the black lines in the dispersion diagrams in
Fig.~\ref{gaps_fig1} cross the value of $k_\|$ inspected in cases
(a)-(c). The blue and red dash-dot vertical lines in
Fig.~\ref{gaps_fig4} indicate where the dips in extinction
associated with the Rayleigh anomalies are seen in the experiment,
which are also the eigenfrequencies used in the oscillator model. We
see in Fig.~\ref{gaps_fig4} that as the nanorod widens and the gap
opens there is an increased frequency deviation of the extinction
dip with respect to the corresponding theoretically predicted
Rayleigh anomalies. Figure~\ref{gaps_fig2} displays the same
phenomenon for normal incidence light, but the effect is much less
pronounced than at the slowdown point. A small shift between the
diffraction edge and the associated extinction dip was also observed
by Augi\'{e} and Barnes in the normal incidence extinction spectra
of various metallic nanoparticle arrays~\cite{Auguie&Barnes08}, but
its origin was not discussed.  It is crucial to realize that when
deriving the conditions for which a diffracted order propagates
grazing to the surface of the grating, i.e. the Rayleigh anomaly
equation yielding the black lines in Fig.~\ref{gaps_fig1} and the
dashed lines in Figs.~\ref{gaps_fig2} and~\ref{gaps_fig4}, the form
factor of the grating is not considered. The derivation follows from
equating the incoming and outgoing wave vectors, the former
including the momentum added by the grating. However, this first
order analysis neglects any changes in momentum that could arise
from the dimensions of the particles constituting the grating. Since
the dimensions of the individual particles determine their
polarizability, i.e., the frequency and damping of the associated
LSPR,  we propose that the small shift in frequency of the
extinction dip with respect to the corresponding theoretically
predicted Rayleigh anomaly may be rooted in the coupling strength of
LSPRs to diffracted orders. Indeed, we observe in
Figs.~\ref{gaps_fig1},~\ref{gaps_fig2} and~\ref{gaps_fig4} that for
the widest nanorods, where the coupling strength of the LSPR to the
($\pm$ 1,0) Rayleigh anomalies is highest, the aforementioned shift
is also highest.

\begin{table*}
\caption{Coupling and damping frequencies used in equation 1 to
reproduce the spectra shown in Figure 4. In the entries for which a
minimum estimate is given, the value in parenthesis is the value
used in the model yielding the spectra in Figure 8. All quantities
are given in units of mrad/nm}
\begin{ruledtabular}
\begin{tabular}{l|ccccccc}
 & $\omega_2$    &$\omega_3$ & $\Omega_{12}$  & $\Omega_{13}$  & $\Omega_{23}$  & $\gamma_{2}$ & $\gamma_{3}$\\
 \hline
$k_\|$=0.13  &   7.31  &   7.01  &   3.0 $\pm$ 0.1  &  $<1.0$ (0.4)  & $1.0 \pm 0.1$   & $<0.02$ (0.008)  &    $0.08 \pm 0.03$    \\
$k_\|$=0.18  &   7.34  &   6.93  &   3.2 $\pm$ 0.1  &  $<1.0$ (0.5)  & $1.2 \pm 0.1$   & $<0.02$ (0.008)  &    $0.08 \pm 0.02$    \\
$k_\|$=0.35  &   7.40  &   6.69  &   3.5 $\pm$ 0.1  &  $<1.0$ (0.8)  & $1.5 \pm 0.1$   & $<0.02$ (0.008)  &    $0.08 \pm 0.01$     \\
\end{tabular}
\end{ruledtabular}
\end{table*}
\begin{figure}
\centerline{\scalebox{0.55}{\includegraphics{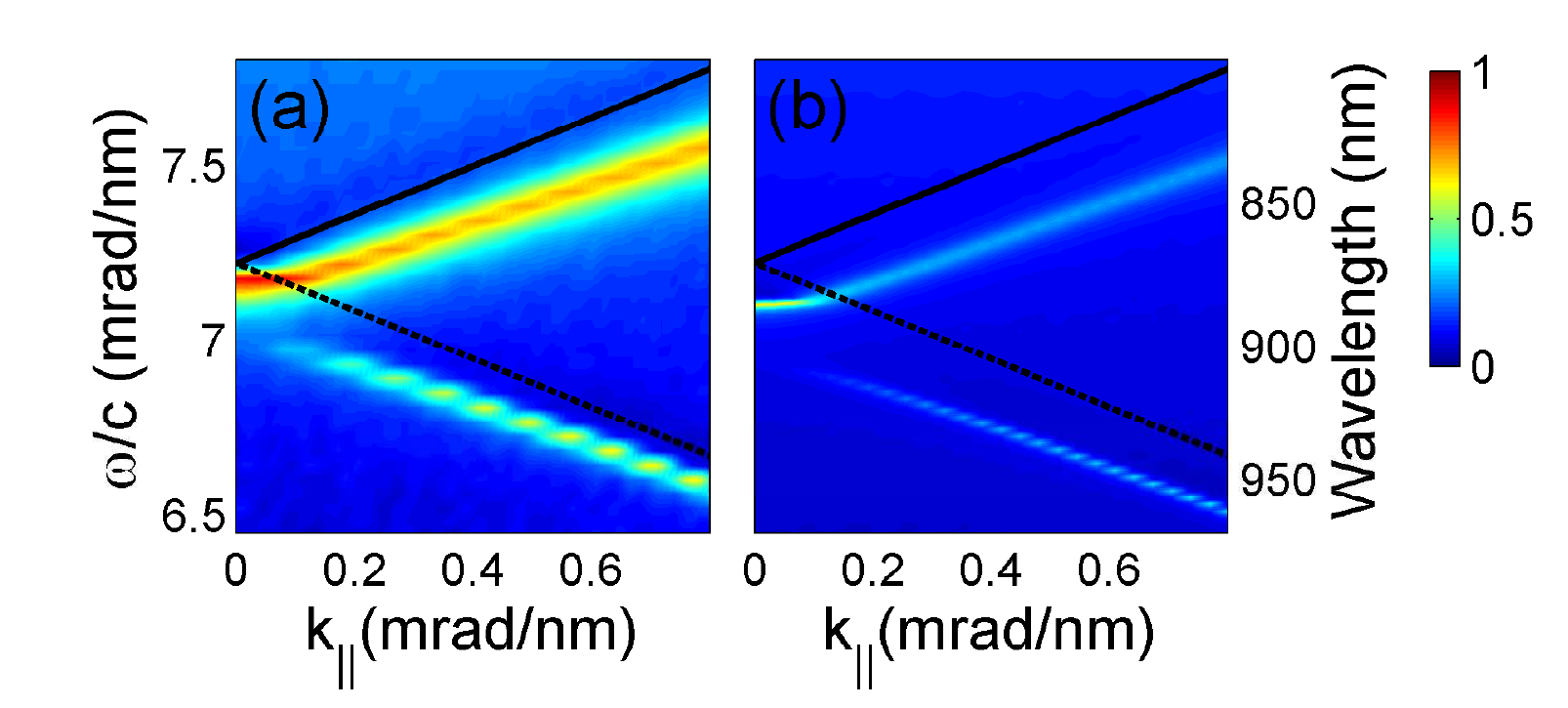}}}
\caption{ (Color online) Figure (a) is a zoom into the stop-gap
displayed in the measurements of Figure 1(a). Figure (b) shows FDTD
simulations results for an array of  comparable antennas. The
extinction is displayed by the same color scale for both
measurements and simulations. The black solid and dashed lines
represent the (+1,0) and (-1,0) Rayleigh anomalies,
respectively.}\label{gaps_fig5}
\end{figure}

 In order to investigate the properties of the gap for a broader
range of nanorod widths, we have performed Finite Difference in Time
Domain (FDTD) simulations with an in-house developed software. We
have calculated dispersion relations in extinction for arrays  with
lattice constants $a_x = 600$ nm and $a_y = 300$ nm, and rods of
dimensions $L = 450$ nm, $h=40$ nm. The width of the rods was varied
between  $w =20$ nm and $w=280$ nm, in steps of 20 nm. Although the
three experimental arrays displayed a small change in the length of
the nanorods, the dominant contribution to the properties of the gap
stems from the width of the nanorod (within the range of length
variation). This assumption is based on the fact that the coupled
surface modes herein discussed arise for light polarized parallel to
the width of the nanorods, and the associated LSPR red-shifts as the
width increases, which is the expected behavior due to the
depolarization field along this dimension~\cite{Maier}. We validate
the simulation results in Fig.~\ref{gaps_fig5}, where we compare the
measured and calculated dispersion relations near the gap for an
array of gold nanorods of 80 nm width in both cases. From the
dimensions of the three experimental arrays previously given, it can
be recognized that the nanorods in the simulations are about $7\% $
shorter in length. Nevertheless, a good qualitative agreement is
observed between measurements and simulations.

\begin{figure}[b]
\centerline{\scalebox{0.55}{\includegraphics{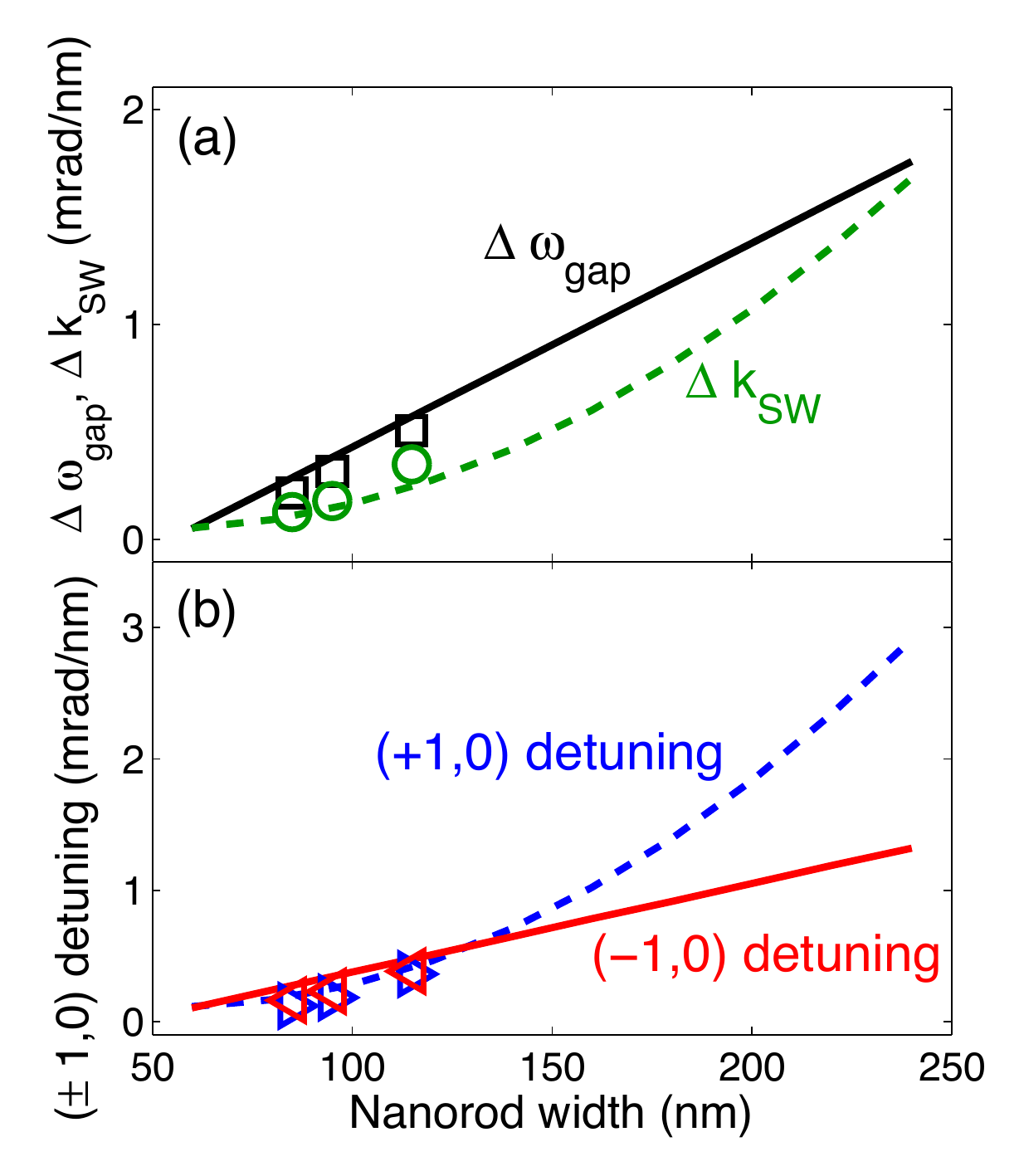}}}
\caption{ (Color online) Figure (a) shows the frequency width of the
stop-gap, $\Delta \omega_{gap}$, and the $k_\|$ width of the
standing waves in the high frequency SLR band, $\Delta k_{SW}$. The
black open squares and solid line represent measurements and FDTD
simulations results, respectively, for $\Delta \omega_{gap}$.  The
green open circles and dashed curve represent measurements and FDTD
simulations results, respectively, for $\Delta k_{SW}$.  Figure (b)
shows the detuning of the ($\pm$1,0) SLRs from their respective
Rayleigh anomalies. The (-1,0) detuning is plotted with red
left-pointing triangles and a solid line, for measurements and FDTD
simulations results, respectively. The (+1,0) detuning is plotted
with blue right-pointing triangles and a dashed curve, for
measurements and FDTD simulations results, respectively. All
quantities in (a) and (b) are plotted as a function of the nanorod
width. }\label{gaps_fig6}
\end{figure}

\begin{figure}[b]
\centerline{\scalebox{0.55}{\includegraphics{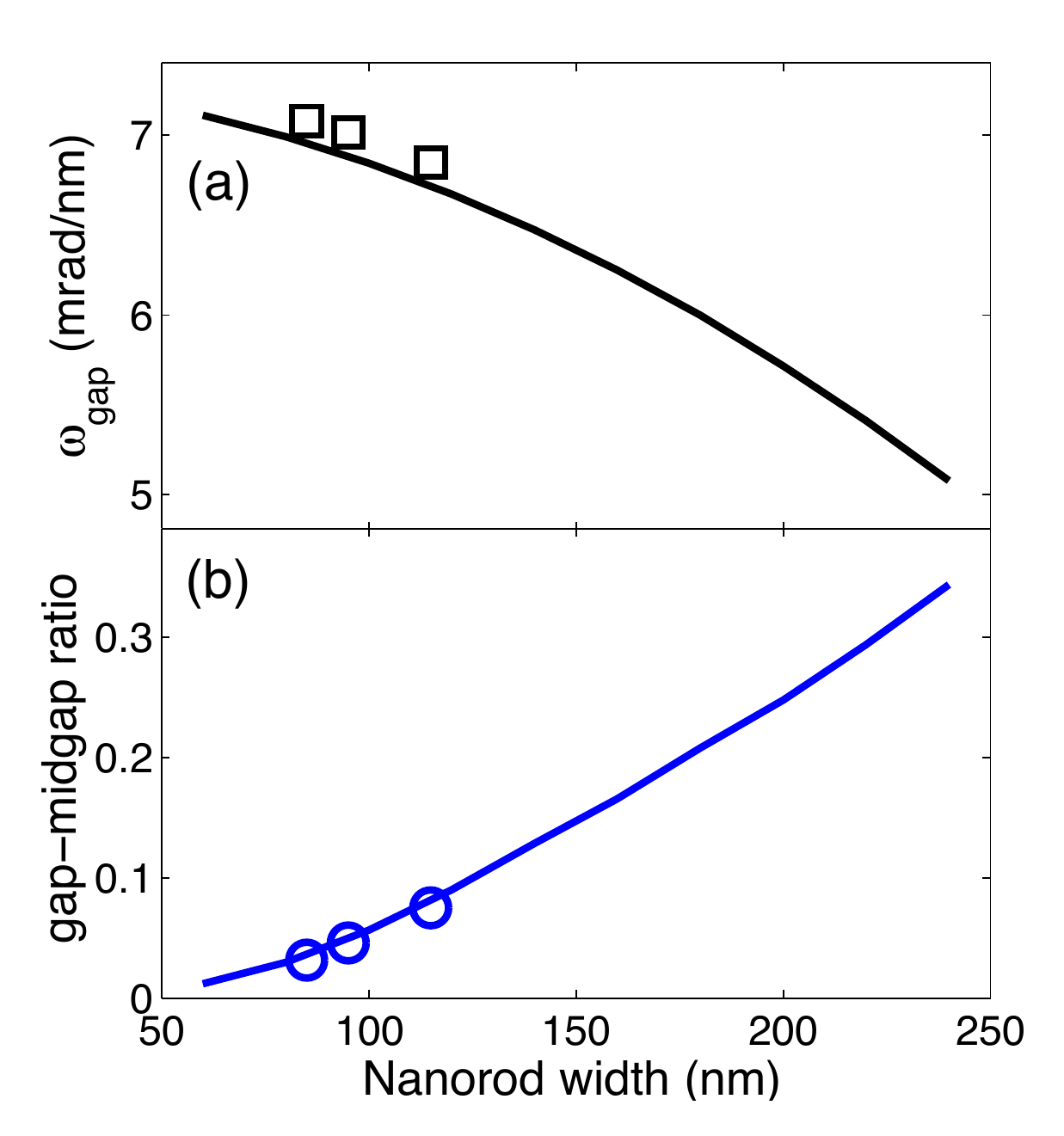}}}
\caption{ (Color online) Figure (a) shows the central frequency of
the stop-gap, $\omega_{gap}$, with black open squares representing
measurements and a black solid curve  representing FDTD simulation
results. Figure (b) shows the gap-midgap ratio, with blue open
circles representing measurements and a blue solid curve
representing FDTD simulations results. All quantities in (a) and (b)
are plotted as a function of the nanorod width. }\label{gaps_fig7}
\end{figure}

In Fig.~\ref{gaps_fig6}(a) we plot the frequency width of the
stop-gap, $\Delta \omega_{gap}$, and the $k_\|$ width of the
standing waves in the high frequency SLR band, $\Delta k_{SW}$, both
as function of the width of the nanorods. $\Delta \omega_{gap}$ is
taken to be the frequency difference between the slowdown point and
the (-1,0) SLR, both evaluated at the $k_\|$  value of the slowdown
point. $\Delta k_{SW}$ is equal to the value of $k_\|$ at the
slowdown point. In Fig.~\ref{gaps_fig6}(b) we plot the detuning of
the ($\pm 1, 0$) SLRs from their respective theoretically predicted
Rayleigh anomalies, also as a function of the nanorod width. The
open data points are taken from the measurements of the three
arrays, and the solid and dashed curves result from the simulations.
A central finding is that both $\Delta \omega_{gap}$ and the (-1,0)
detuning are linearly increasing functions of the nanorod width,
while $\Delta k_{SW}$ and the (+1,0) detuning are both quadratically
increasing functions of the nanorod width. We therefore find the
frequency width of the gap to be correlated with the (-1,0)
detuning, and the $k_\|$ width of the standing waves to be
correlated with the (+1,0) detuning. In terms of the coupled
oscillator model, this can be interpreted as follows: the gap opens
with a linear increase in $\Omega_{13}$ (coupling of LSPR to the
(-1,0) order), whereas the in-plane momentum width of the standing
waves broadens with a quadratic increase in $\Omega_{12}$ (coupling
of LSPR to the (+1,0) order). The latter dependance is the reason
for which  $\Omega_{12}$ was fitted with a quadratic function in the
normal incidence spectra, i.e., the dashed curve in
Fig.~\ref{gaps_fig3}(c).

 Further quantifying the properties of the gap, in Fig.~\ref{gaps_fig7}(a) we
plot the central frequency of the stop-gap, $\omega_{gap}$, which
falls quadratically for increasing nanorod width. In
Fig.~\ref{gaps_fig7}(b) we plot the gap-midgap ratio, $\Delta
\omega_{gap}/ \omega_{gap}$ which rises quadratically for increasing
nanorod width. Figure~\ref{gaps_fig7}(a)shows that for nanorods
wider than those considered in the experiments, the central
frequency of the gap falls into the infrared part of the spectrum.
Limited by the spectral response of our acquisition system (based on
a silicon detector), we are unable to perform measurements for the
wider nanorods. Nevertheless, in light of the good agreement between
measurements and simulations we believe Figs.~\ref{gaps_fig6}(a) and
~\ref{gaps_fig7} accurately encompass all properties of the gap as
function of the nanorod width, and provide a suitable recipe upon
which stop gaps in plasmonic crystals may be opened and tuned.
Comparing to previous work, where it was found that the gap's width
is a linearly increasing function of the modulation amplitude in
metallic sinusoidal gratings~\cite{Barnes96}, we have also found a
linear dependance of the gap's width on a single structural factor.
However, the connection between these two works is not an obvious
one, since whereas Barnes and co-workers have varied a dimension out
of the plane of propagation, we have investigated structures of
equal height but variable dimensions in the plane of propagation.
From a fabrication point of view, a precise in plane structuring of
plasmonic crystals may offer a higher degree of versatility in how
stop-gaps may be tuned and at a greater ease.

\begin{figure}
\centerline{\scalebox{0.5}{\includegraphics{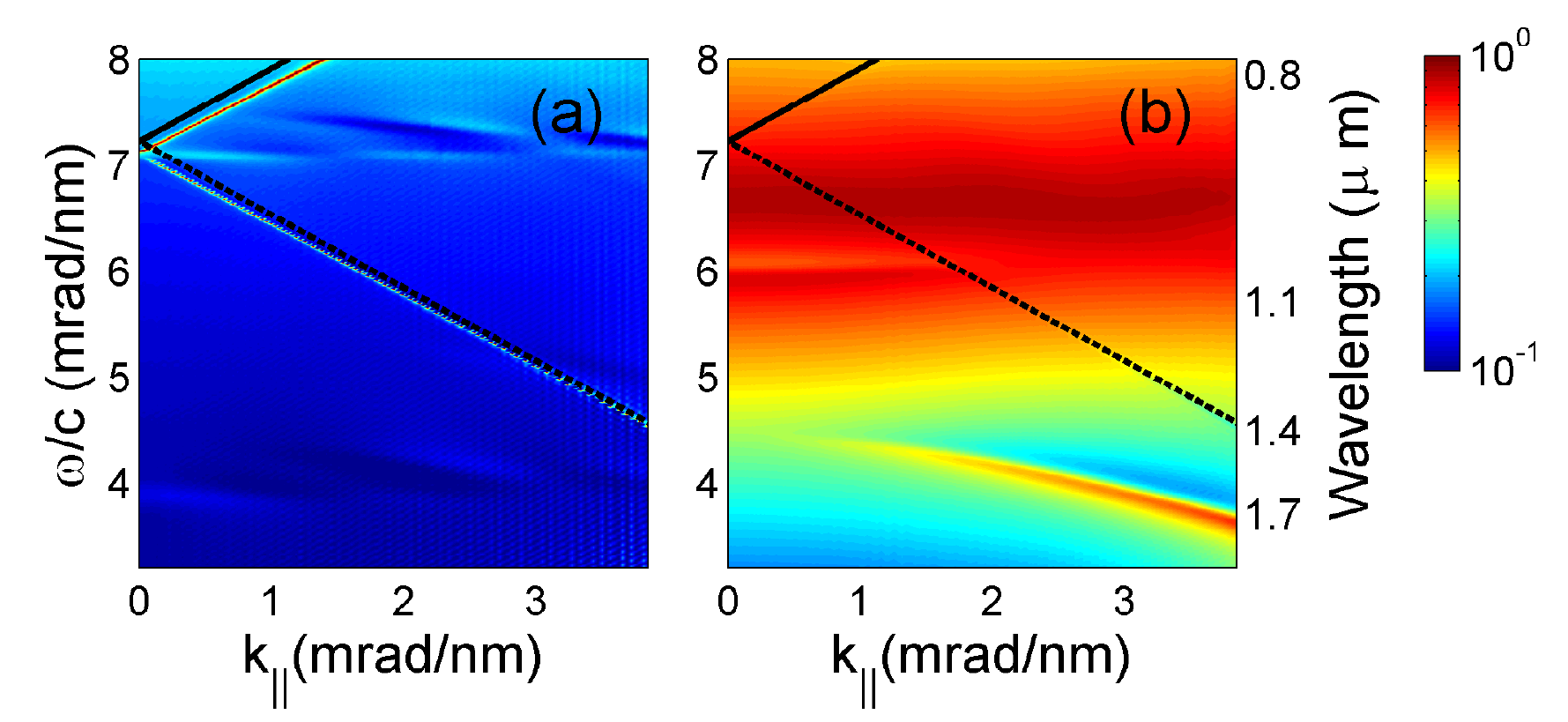}}} \caption{
(Color online) FDTD simulations results for the extinction of an
array of gold nanorods of width (a) 60 nm and (b) 240 nm. Both
arrays have nanorods of length = 450 nm, height = 40 nm, in a
lattice with constants $a_x = 600$ nm and $a_y = 300$ nm. The black
solid and dashed lines indicate the (+1,0) and (-1,0) Rayleigh
anomalies, respectively. The extinction is displayed by a
logarithmic color scale common to both graphs. }\label{gaps_fig8}
\end{figure}

We now discuss the validity range of the previously discussed SLR
properties on the nanorod width. Figure~\ref{gaps_fig8} shows the
extinction of two arrays with lattice constants $a_x = 600$ nm, $a_y
= 300$ nm, nanorods of length$= 450$ nm and height $= 40$ nm, and
embedded in a homogeneous environment of n=1.45; these conditions
are the same as in the experiments presented in
Fig.~\ref{gaps_fig1}. The width of the nanorods in
Fig.~\ref{gaps_fig8} is (a) 60 nm, and (b) 240 nm, which correspond
to the extremes of the curves plotted in Figs.~\ref{gaps_fig6}
and~\ref{gaps_fig7}. Due to the very different extinction efficiency
of the two arrays in the spectral region of interest, the extinction
is displayed by a logarithmic color scale common to both graphs,
which allows quantitative comparison of the spectra. The low
extinction and remarkably narrow resonances seen in
Fig.~\ref{gaps_fig8}(a) can be explained in terms of the coupled
oscillator model as a consequence of the low coupling strength of
LSPRs to diffracted orders. This point can be inferred from the
quantities given in Table 1, where all coupling frequencies
$\Omega_{jk}$ are seen to decrease as the nanorod width decreases.
Lower $\Omega_{12}$ and $\Omega_{13}$ therefore translate into a
lower coupling strength of LSPRs to diffracted orders, whereas a
lower $\Omega_{23}$ translates into a lower coupling between the
(+1,0) and (-1,0) SLRs and therefore a smaller gap. As the width of
the nanorods increases and the $\Omega_{jk}$ terms increase, the
extinction at the SLRs also increases, the resonance linewidth
broadens, and the frequency gap widens, eventually reaching the case
displayed in Fig.~\ref{gaps_fig8}(b) for a nanorod width of $240$
nm. Figure~\ref{gaps_fig8}(b) displays a dipolar LSPR near 6.8
mrad/nm, which is lower in frequency than the diffraction edge at
normal incidence. The (+1,0) SLR can still be recognized in the
spectrum from the non-dispersive feature near 6.0 mrad/nm (in the
red tail of the LSPR). However, for wider nanorods the LSPR shifts
to lower frequencies, thereby making the (+1,0) SLR
indistinguishable from the LSPR. As the energy of the dipolar LSPR
becomes substantially lower that the ($\pm1,0$) diffraction orders,
diffractive coupling of dipolar LSPRs becomes very weak, and the
properties of coupled SLRs leading to the opening of the stop-gap
significantly deviate.

In conclusion, we have shown how the radiative coupling strength of
localized surface plasmons to diffracted orders in periodic arrays
of nanorods can be tuned. This tuning is achieved experimentally by
modifying the width of the nanorods, and results in the opening of
frequency stop-gaps whose properties are therefore tunable. A
quadratic dependence of both the frequency width of the stop-gap and
the (-1,0) SLR detuning was found on the nanorod width. A linear
dependence of both the in-plane momentum width of the standing waves
in the high-frequency SLR band and the (+1,0) SLR detuning was found
on the nanorod width. In light of a coupled oscillator analog to the
plasmonic crystal, we have associated these two correlations with
the coupling strength of localized surface plasmons to the (-1,0)
and (+1,0) diffracted orders. Supporting experiments with numerical
simulations for a wider set of nanorod widths, we have analyzed the
properties of the gap and discussed the limiting cases where the
scaling laws we present deviate. Although we have only considered
nanorod arrays in this work, similar results are expected to hold in
periodic arrays of particles with different geometries whereby
diffractive coupling of localized surface plasmons is possible. Our
results therefore pave the road towards nanoscale light manipulation
in 2D plasmonic crystals, since stop-gaps allow to selectively
enhance or suppress light-matter interactions.

We thank M. Verschuuren for assistance in the fabrication of the
samples, and G. Lozano for fruitful discussions. This work was
supported by the Netherlands Foundation Fundamenteel Onderzoek der
Materie (FOM) and the Nederlandse Organisatie voor Wetenschappelijk
Onderzoek (NWO), and is part of an industrial partnership program
between Philips and FOM.


\begin{thebibliography}{21}
\expandafter\ifx\csname
natexlab\endcsname\relax\def\natexlab#1{#1}\fi
\expandafter\ifx\csname bibnamefont\endcsname\relax
  \def\bibnamefont#1{#1}\fi
\expandafter\ifx\csname bibfnamefont\endcsname\relax
  \def\bibfnamefont#1{#1}\fi
\expandafter\ifx\csname citenamefont\endcsname\relax
  \def\citenamefont#1{#1}\fi
\expandafter\ifx\csname url\endcsname\relax
  \def\url#1{\texttt{#1}}\fi
\expandafter\ifx\csname urlprefix\endcsname\relax\def\urlprefix{URL
}\fi \providecommand{\bibinfo}[2]{#2}
\providecommand{\eprint}[2][]{\url{#2}}

\bibitem[{\citenamefont{Joannopoulos et~al.}(2008)\citenamefont{Joannopoulos,
  Johnson, Winn, and Meade}}]{Joann}
\bibinfo{author}{\bibfnamefont{J.~D.} \bibnamefont{Joannopoulos}},
  \bibinfo{author}{\bibfnamefont{S.~G.} \bibnamefont{Johnson}},
  \bibinfo{author}{\bibfnamefont{J.~N.} \bibnamefont{Winn}}, \bibnamefont{and}
  \bibinfo{author}{\bibfnamefont{R.~D.} \bibnamefont{Meade}},
  \emph{\bibinfo{title}{Photonic Crystals: Molding the Flow of Light}}
  (\bibinfo{publisher}{Princeton University Press}, \bibinfo{address}{New
  Jersey, USA}, \bibinfo{year}{2008}), \bibinfo{edition}{2nd} ed.

\bibitem[{\citenamefont{Kitson et~al.}(1996)\citenamefont{Kitson, Barnes, and
  Sambles}}]{Barnes96b}
\bibinfo{author}{\bibfnamefont{S.~C.} \bibnamefont{Kitson}},
  \bibinfo{author}{\bibfnamefont{W.~L.} \bibnamefont{Barnes}},
  \bibnamefont{and} \bibinfo{author}{\bibfnamefont{J.~R.}
  \bibnamefont{Sambles}}, \bibinfo{journal}{Phys. Rev. Lett.}
  \textbf{\bibinfo{volume}{77}}, \bibinfo{pages}{2670} (\bibinfo{year}{1996}).

\bibitem[{\citenamefont{Barnes et~al.}(1996)\citenamefont{Barnes, Preist,
  Kitson, and Sambles}}]{Barnes96}
\bibinfo{author}{\bibfnamefont{W.~L.} \bibnamefont{Barnes}},
  \bibinfo{author}{\bibfnamefont{T.~W.} \bibnamefont{Preist}},
  \bibinfo{author}{\bibfnamefont{S.~C.} \bibnamefont{Kitson}},
  \bibnamefont{and} \bibinfo{author}{\bibfnamefont{J.~R.}
  \bibnamefont{Sambles}}, \bibinfo{journal}{Phys. Rev. B}
  \textbf{\bibinfo{volume}{54}}, \bibinfo{pages}{6227} (\bibinfo{year}{1996}).

\bibitem[{\citenamefont{Lochbihler}(1994)}]{Lochbihler94}
\bibinfo{author}{\bibfnamefont{H.}~\bibnamefont{Lochbihler}},
  \bibinfo{journal}{Phys. Rev. B} \textbf{\bibinfo{volume}{50}},
  \bibinfo{pages}{4795} (\bibinfo{year}{1994}).

\bibitem[{\citenamefont{Yoon et~al.}(2003)}]{Yoon}
\bibinfo{author}{\bibfnamefont{J.}~\bibnamefont{Yoon}} \bibnamefont{et~al.},
  \bibinfo{journal}{J. Appl. Phys.} \textbf{\bibinfo{volume}{94}},
  \bibinfo{pages}{123} (\bibinfo{year}{2003}).

\bibitem[{\citenamefont{Sauvan et~al.}(2008)\citenamefont{Sauvan, Billaudeau,
  Collin, Bardou, Pardo, Pelouard, and Lalanne}}]{Lalanne08}
\bibinfo{author}{\bibfnamefont{C.}~\bibnamefont{Sauvan}},
  \bibinfo{author}{\bibfnamefont{C.}~\bibnamefont{Billaudeau}},
  \bibinfo{author}{\bibfnamefont{S.}~\bibnamefont{Collin}},
  \bibinfo{author}{\bibfnamefont{N.}~\bibnamefont{Bardou}},
  \bibinfo{author}{\bibfnamefont{F.}~\bibnamefont{Pardo}},
  \bibinfo{author}{\bibfnamefont{J.-L.} \bibnamefont{Pelouard}},
  \bibnamefont{and} \bibinfo{author}{\bibfnamefont{P.}~\bibnamefont{Lalanne}},
  \bibinfo{journal}{Appl. Phys. Lett.} \textbf{\bibinfo{volume}{92}},
  \bibinfo{eid}{011125} (\bibinfo{year}{2008}).

\bibitem[{\citenamefont{Billaudeau et~al.}(2008)\citenamefont{Billaudeau,
  Collin, Sauvan, Bardou, Pardo, and Pelouard}}]{Billaudeau}
\bibinfo{author}{\bibfnamefont{C.}~\bibnamefont{Billaudeau}},
  \bibinfo{author}{\bibfnamefont{S.}~\bibnamefont{Collin}},
  \bibinfo{author}{\bibfnamefont{C.}~\bibnamefont{Sauvan}},
  \bibinfo{author}{\bibfnamefont{N.}~\bibnamefont{Bardou}},
  \bibinfo{author}{\bibfnamefont{F.}~\bibnamefont{Pardo}}, \bibnamefont{and}
  \bibinfo{author}{\bibfnamefont{J.-L.} \bibnamefont{Pelouard}},
  \bibinfo{journal}{Opt. Lett.} \textbf{\bibinfo{volume}{33}},
  \bibinfo{pages}{165} (\bibinfo{year}{2008}).

\bibitem[{\citenamefont{Ropers et~al.}(2005)\citenamefont{Ropers, Park,
  Stibenz, Steinmeyer, Kim, Kim, and Lienau}}]{Ropers}
\bibinfo{author}{\bibfnamefont{C.}~\bibnamefont{Ropers}},
  \bibinfo{author}{\bibfnamefont{D.~J.} \bibnamefont{Park}},
  \bibinfo{author}{\bibfnamefont{G.}~\bibnamefont{Stibenz}},
  \bibinfo{author}{\bibfnamefont{G.}~\bibnamefont{Steinmeyer}},
  \bibinfo{author}{\bibfnamefont{J.}~\bibnamefont{Kim}},
  \bibinfo{author}{\bibfnamefont{D.~S.} \bibnamefont{Kim}}, \bibnamefont{and}
  \bibinfo{author}{\bibfnamefont{C.}~\bibnamefont{Lienau}},
  \bibinfo{journal}{Phys. Rev. Lett.} \textbf{\bibinfo{volume}{94}},
  \bibinfo{pages}{113901} (\bibinfo{year}{2005}).

\bibitem[{\citenamefont{Ghoshal et~al.}(2009)\citenamefont{Ghoshal, Divliansky,
  and Kik}}]{Ghoshal&Kik09}
\bibinfo{author}{\bibfnamefont{A.}~\bibnamefont{Ghoshal}},
  \bibinfo{author}{\bibfnamefont{I.}~\bibnamefont{Divliansky}},
  \bibnamefont{and} \bibinfo{author}{\bibfnamefont{P.~G.} \bibnamefont{Kik}},
  \bibinfo{journal}{Appl. Phys. Lett.} \textbf{\bibinfo{volume}{94}},
  \bibinfo{eid}{171108} (\bibinfo{year}{2009}).

\bibitem[{\citenamefont{Augui\'e and Barnes}(2008)}]{Auguie&Barnes08}
\bibinfo{author}{\bibfnamefont{B.}~\bibnamefont{Augui\'e}} \bibnamefont{and}
  \bibinfo{author}{\bibfnamefont{W.~L.} \bibnamefont{Barnes}},
  \bibinfo{journal}{Phys. Rev. Lett.} \textbf{\bibinfo{volume}{101}},
  \bibinfo{pages}{143902} (\bibinfo{year}{2008}).

\bibitem[{\citenamefont{Chu et~al.}(2008)\citenamefont{Chu, Schonbrun, Yang,
  and Crozier}}]{Crozier}
\bibinfo{author}{\bibfnamefont{Y.}~\bibnamefont{Chu}},
  \bibinfo{author}{\bibfnamefont{E.}~\bibnamefont{Schonbrun}},
  \bibinfo{author}{\bibfnamefont{T.}~\bibnamefont{Yang}}, \bibnamefont{and}
  \bibinfo{author}{\bibfnamefont{K.~B.} \bibnamefont{Crozier}},
  \bibinfo{journal}{Appl. Phys. Lett.} \textbf{\bibinfo{volume}{93}},
  \bibinfo{pages}{181108} (\bibinfo{year}{2008}).

\bibitem[{\citenamefont{Kravets et~al.}(2008)\citenamefont{Kravets, Schedin,
  and Grigorenko}}]{Kravets08}
\bibinfo{author}{\bibfnamefont{V.~G.} \bibnamefont{Kravets}},
  \bibinfo{author}{\bibfnamefont{F.}~\bibnamefont{Schedin}}, \bibnamefont{and}
  \bibinfo{author}{\bibfnamefont{A.~N.} \bibnamefont{Grigorenko}},
  \bibinfo{journal}{Phys. Rev. Lett.} \textbf{\bibinfo{volume}{101}},
  \bibinfo{pages}{087403} (\bibinfo{year}{2008}).

\bibitem[{\citenamefont{Vecchi et~al.}(2009)\citenamefont{Vecchi, Giannini, and
  G\'omez~Rivas}}]{Vecchi09b}
\bibinfo{author}{\bibfnamefont{G.}~\bibnamefont{Vecchi}},
  \bibinfo{author}{\bibfnamefont{V.}~\bibnamefont{Giannini}}, \bibnamefont{and}
  \bibinfo{author}{\bibfnamefont{J.}~\bibnamefont{G\'omez~Rivas}},
  \bibinfo{journal}{Phys. Rev. B} \textbf{\bibinfo{volume}{80}},
  \bibinfo{pages}{201401} (\bibinfo{year}{2009}).

\bibitem[{\citenamefont{Bitzer et~al.}(2009)\citenamefont{Bitzer, Wallauer,
  Merbold, Helm, Feurer, and Walther}}]{Bitzer}
\bibinfo{author}{\bibfnamefont{A.}~\bibnamefont{Bitzer}},
  \bibinfo{author}{\bibfnamefont{J.}~\bibnamefont{Wallauer}},
  \bibinfo{author}{\bibfnamefont{H.}~\bibnamefont{Merbold}},
  \bibinfo{author}{\bibfnamefont{H.}~\bibnamefont{Helm}},
  \bibinfo{author}{\bibfnamefont{T.}~\bibnamefont{Feurer}}, \bibnamefont{and}
  \bibinfo{author}{\bibfnamefont{M.}~\bibnamefont{Walther}},
  \bibinfo{journal}{Opt. Exp.} \textbf{\bibinfo{volume}{17}},
  \bibinfo{pages}{22108} (\bibinfo{year}{2009}).

\bibitem[{\citenamefont{Rodriguez et~al.}(2011)\citenamefont{Rodriguez, Abass,
  Maes, Janssen, Vecchi, and G\'{o}mez~Rivas}}]{Rodriguez11}
\bibinfo{author}{\bibfnamefont{S.~R.~K.} \bibnamefont{Rodriguez}},
  \bibinfo{author}{\bibfnamefont{A.}~\bibnamefont{Abass}},
  \bibinfo{author}{\bibfnamefont{B.}~\bibnamefont{Maes}},
  \bibinfo{author}{\bibfnamefont{O.~T.~A.} \bibnamefont{Janssen}},
  \bibinfo{author}{\bibfnamefont{G.}~\bibnamefont{Vecchi}}, \bibnamefont{and}
  \bibinfo{author}{\bibfnamefont{J.}~\bibnamefont{G\'{o}mez~Rivas}}
  (\bibinfo{year}{2011}), \eprint{arXiv:1108.1620v1}.

\bibitem[{\citenamefont{Verschuuren}(2010)}]{scil}
\bibinfo{author}{\bibfnamefont{M.~A.} \bibnamefont{Verschuuren}}, Ph.D. thesis,
  \bibinfo{school}{Utrecht University} (\bibinfo{year}{2010}).

\bibitem[{\citenamefont{Giannini et~al.}(2010)\citenamefont{Giannini, Vecchi,
  and G\'omez~Rivas}}]{Giannini10}
\bibinfo{author}{\bibfnamefont{V.}~\bibnamefont{Giannini}},
  \bibinfo{author}{\bibfnamefont{G.}~\bibnamefont{Vecchi}}, \bibnamefont{and}
  \bibinfo{author}{\bibfnamefont{J.}~\bibnamefont{G\'omez~Rivas}},
  \bibinfo{journal}{Phys. Rev. Lett.} \textbf{\bibinfo{volume}{105}},
  \bibinfo{pages}{266801} (\bibinfo{year}{2010}).

\bibitem[{\citenamefont{Garrido~Alzar et~al.}(2002)\citenamefont{Garrido~Alzar,
  Martinez, and Nussenzveig}}]{Alzar05}
\bibinfo{author}{\bibfnamefont{C.~L.} \bibnamefont{Garrido~Alzar}},
  \bibinfo{author}{\bibfnamefont{M.~A.~G.} \bibnamefont{Martinez}},
  \bibnamefont{and}
  \bibinfo{author}{\bibfnamefont{P.}~\bibnamefont{Nussenzveig}},
  \bibinfo{journal}{Am. J. Phys.} \textbf{\bibinfo{volume}{70}},
  \bibinfo{pages}{37} (\bibinfo{year}{2002}).

\bibitem[{\citenamefont{Mukherjee et~al.}(2010)\citenamefont{Mukherjee,
  Sobhani, Lassiter, Bardhan, Nordlander, and Halas}}]{Halas10}
\bibinfo{author}{\bibfnamefont{S.}~\bibnamefont{Mukherjee}},
  \bibinfo{author}{\bibfnamefont{H.}~\bibnamefont{Sobhani}},
  \bibinfo{author}{\bibfnamefont{J.~B.} \bibnamefont{Lassiter}},
  \bibinfo{author}{\bibfnamefont{R.}~\bibnamefont{Bardhan}},
  \bibinfo{author}{\bibfnamefont{P.}~\bibnamefont{Nordlander}},
  \bibnamefont{and} \bibinfo{author}{\bibfnamefont{N.~J.} \bibnamefont{Halas}},
  \bibinfo{journal}{Nano Lett.} \textbf{\bibinfo{volume}{10}},
  \bibinfo{pages}{2694} (\bibinfo{year}{2010}).

\bibitem[{\citenamefont{Maier}(2007)}]{Maier}
\bibinfo{author}{\bibfnamefont{S.~A.} \bibnamefont{Maier}},
  \emph{\bibinfo{title}{Plasmonics: Fundamentals and Applications}}
  (\bibinfo{publisher}{Springer}, \bibinfo{address}{New York, USA},
  \bibinfo{year}{2007}).

\bibitem[{\citenamefont{Offermans et~al.}(2011)\citenamefont{Offermans,
  Schaafsma, Rodriguez, Zhang, Crego-Calama, Brongersma, and
  G\'{o}mez~Rivas}}]{Offermans}
\bibinfo{author}{\bibfnamefont{P.}~\bibnamefont{Offermans}},
  \bibinfo{author}{\bibfnamefont{M.~C.} \bibnamefont{Schaafsma}},
  \bibinfo{author}{\bibfnamefont{S.~R.~K.} \bibnamefont{Rodriguez}},
  \bibinfo{author}{\bibfnamefont{Y.}~\bibnamefont{Zhang}},
  \bibinfo{author}{\bibfnamefont{M.}~\bibnamefont{Crego-Calama}},
  \bibinfo{author}{\bibfnamefont{S.~H.} \bibnamefont{Brongersma}},
  \bibnamefont{and}
  \bibinfo{author}{\bibfnamefont{J.}~\bibnamefont{G\'{o}mez~Rivas}},
  \bibinfo{journal}{ACS Nano} \textbf{\bibinfo{volume}{5}},
  \bibinfo{pages}{5151} (\bibinfo{year}{2011}).

\end{thebibliography}
\end{document}